\def\gsim{\;\raisebox{-.4ex}{\rlap{$\sim$}} \raisebox{.4ex}{$>$}\;}

\documentstyle[12pt]{article}
\begin{document}

\begin{titlepage}

\begin{flushright}
FERMILAB-Pub-96/435-T\\
hep-ph9612265\\
{\small December 1996}
\end{flushright}

\vspace{1cm}

\begin{center}
{\Large\bf A Way to Measure Very Large $\Delta m$ for $B_s$ Mesons }

\vspace{1cm}

{\large Yakov Azimov${}^{(a,b)},\;\;$ Isard Dunietz${}^{(a)}$}\\

\vspace{0.4cm}

{\it${}^{(a)}$Fermi National Accelerator Laboratory, P.~O.~Box 500,\\
Batavia, IL 60510, U.S.A.}\\

{\it ${}^{(b)}$Petersburg Nuclear Physics Institute,\\
Gatchina, St.Petersburg, 188350, Russia}\\
\vspace{0.4cm}
\end{center}
\begin{abstract}
While present vertex technology cannot measure $x_s$ much 
beyond 20, the Standard Model accomodates significantly 
larger $x_s$ values. This note presents a method to 
determine very large $x_s$ with present technology. The 
determination is based upon subtle coherence effects
between initial $B_s$ mesons and daughter neutral kaons,
discovered by one of us several years ago. The method 
may be useful also for measuring very small width or mass
differences in mixed neutral $B$ or $D$ mesons.\\
\vspace{0.4cm}

Pacs numbers: 14.40.Nd, 13.25.Hw, 13.25.Es
\end{abstract}
\end{titlepage}
\newpage

The physics of $B-$mesons is an intensively developing 
area of particle physics. It is strongly stimulated by 
the hope to understand the mechanism(s) of $CP-$violation. 
In this respect especially interesting are neutral 
$B_d$ and $B_s$ mesons~\cite{BH}. But to use them as a 
testing ground for $CP-$studies one should know, to good 
precision and in great detail, the basic properties of 
neutral $B-$mesons themselves. Ways to attain this goal 
are rather well investigated for  $B_d$. The situation 
for $B_s$ appears more confused.

The difficulty is mainly a consequence of the expected 
large mass difference between the two eigenstates of 
$B_s$ (analogous to $K_S$ and $K_L$ for kaons). The most 
direct traditional method~\cite{ALEPH} to measure this 
mass difference $\Delta m_s$  is to find and trace the 
$(\Delta m_st)$-oscillations of flavor-specific 
$B_s-$decays which, in addition, must be flavor-tagged. 
Ref.~\cite{Dig} pointed out that the $B_s-$decay modes 
need not be flavor-specific and that the value of 
$\Delta m_s$ could be extracted from time-dependences 
of such flavor-tagged decays as, e.g., $B_s\rightarrow 
J/\psi\,\phi$. Both methods require superb vertex 
detectors to resolve very rapid oscillations.

At present, LEP experiments~\cite{ALEPH} report a limit 
on the $B_s$ mass difference  
$$|x_s| = \frac{|\Delta m_s|}{\Gamma^{(s)}} \gsim 15\,.$$
Current vertex technology is able to resolve only 
marginally larger values of $x_s$ of, say, up to 
about 20~\cite{txt}, while theory accomodates values 
of up to 40 or even higher~[5--7]. The present note 
suggests a novel method which allows the measurement 
of very large $x_s$.
 
If the traditional method is being integrated over time 
to overcome insufficient resolution, the oscillations  still 
give a non-vanishing contribution sensitive to $|x_s|$, 
but the sensitivity drops as $\cal O$$(1/x_s^2)$ 
(for discussion on this point see, e.g., ref.~\cite{AUK}).

In this brief note we show that there exists a triggerable,
though rare, decay of $B_s$  that is more sensitive to $x_s$,
even at large values, and, moreover, sensitive to its sign.
As we will show this allows the experimental identification 
of whether the heavy eigenstate $B_s^H$ is approximately
$CP-$odd (as predicted within the Standard Model) or $CP-$even.

To be concrete, we mean the meson decay
\begin{equation}
B_s (\overline B_s) \rightarrow J/\psi 
\overline K^0 (K^0)\,,
\end{equation}
induced by the quark decay $b \rightarrow c\overline c d$\,.
As demonstrated in refs.~[9--11]\,, decays 
of neutral $B-$mesons producing neutral kaons have some 
unique properties. They are based on the coherence of 
oscillations of the two flavors: beauty before the $B-$meson 
decay, and strangeness after it. When studying decays 
of this daughter kaon,  such double coherence 
generates various unusual effects and gives a better 
insight into details of mixing and $CP-$violation~[9--12].

For simplicity we begin with a comparison of the decay 
(1) and the similar decay
\begin{equation}
B_d (\overline B_d) \rightarrow J/\psi K^0
(\overline K^0 )\,,
\end{equation}
induced by the more intensive quark decay $b \rightarrow 
c\overline c s$\,. Of course, they differ in intensity
due to different CKM matrix elements (the ratio of amplitudes 
is approximately $\tan\theta_C $). But there is another 
difference as well. Decay (2) admits only the transitions 
\begin{equation}
B_d\rightarrow K^0\,,\;\;\;\;\;\;
\overline B_d \rightarrow \overline K^0\,,
\end{equation}
while the only transitions in the decay (1) are
\begin{equation}
B_s \rightarrow \overline K^0\,,\;\;\;\;\;\;
\overline B_s \rightarrow  K^0\,.
\end{equation}
So, in the decay (2), transitions (3) provide equality of final 
strangeness $S_f$ and initial beauty $B_i$. Transitions (4) for 
the decay (1) make $S_f$ and $B_i$ have opposite signs. But in 
both cases $S_f$ and $B_i$ are related to each other 
unambiguously. It is just this fact that makes the final state 
(before the kaon decay) be coherent to the initial one for both 
decays (1) and (2). As a result, evolution of the produced kaon 
retains~\,``memory" of the initial $B-$evolution. And after the 
daughter kaon also decays, distributions of this secondary decay 
reveal information on evolution and decay properties of the initial 
$B-$meson.

Neutral kaons evolve in time orders of magnitude less rapidly than
$B_s-$mesons. Their time-evolution can thus be easily traced at 
present, thereby providing a tool to measure the $B_s - \overline
B_s$ mixing parameters through the above-mentioned coherence effect.

Now we can compare decays (1) and (2) in more detail. We consider
both of them as cascade decays assuming some particular mode for 
the neutral kaons. Most convenient are either semileptonic
decays
\begin{equation} 
K^0(\overline K^0) \rightarrow l^+\nu\pi^- 
(l^-\overline\nu\pi^+)\,,
\end{equation}  
or 2-pion decays
\begin{equation} 
K^0(\overline K^0) \rightarrow \pi^+ \pi^-. 
\end{equation}  
So, every cascade consists of two stages. The first one is 
the evolution and decay (1) or (2) of the initial $B-$meson 
state. The second stage is the evolution and decay (5) or (6) 
of the daughter kaon state. For each stage we use the relevant
proper time in the rest-frame of the $B$ or $K$ respectively. 
Distributions over  time $t_1$ of the first stage of the 
cascade (decay time of $B$) and time $t_2$ of the second stage 
(decay time of kaon) are strongly correlated in both cases. 
The form of the correlation depends on the kaon decay mode and 
on one more parameter. For the decay (2) it is 
\begin{equation}
\lambda_d = \frac{1 - \epsilon_d}{1 + \epsilon_d}\cdot 
\frac{1 +\epsilon_K}{1 - \epsilon_K}\cdot 
\frac{\overline a_d}{a_d}\,
\end{equation}
(see ref.~\cite{A3}, with minor differences in notations,  
e.g., $\lambda$ instead of $\lambda_d$; in what follows, for 
particular expressions we refer to the paper~\cite{A3},\,which 
supersedes papers~[9, 10]). Amplitudes $a_d$ 
and $\overline a_d$ correspond to the transitions 
$$B_d\rightarrow J/\psi K^0\,,\,\,\,\,\,\, 
\overline B_d\rightarrow J/\psi \overline K^0\,. $$
Similar calculations for the decay (1) produce the analogous 
parameter
\begin{equation}
\lambda_s = \frac{1 - \epsilon_s}{1 + \epsilon_s}\cdot 
\frac{1 - \epsilon_K}{1 + \epsilon_K}\cdot 
\frac{\overline a_s}{a_s}\,\,,
\end{equation}
where $a_s\,,\,\,\overline a_s$ are amplitudes for the 
transitions
$$B_s\rightarrow J/\psi \overline K^0\,,\;\;\;\;\;\; 
\overline B_s\rightarrow J/\psi  K^0\,. $$
Note that $B_s(\overline B_s)$ mesons in the decay (1) 
generate the same final state as $\overline B_d(B_d)$ 
in the decay~(2). As a result, time distributions for 
a cascade initiated by the decay (1) may be related 
to distributions for a similar cascade beginning with 
the decay (2) by the simple changes $\lambda_d \rightarrow 
1/\lambda_s,\, a_d \rightarrow \overline a_s,\,  \overline 
a_d \rightarrow  a_s$, which are analogous to changes relating 
initial $B_d$ and $\overline B_d$ states in the decay (2)
(see ref.~\cite{A3}). In such a way  we can easily write 
all the necessary expressions for $B_s-$mesons using the 
corresponding formulas of ref.~\cite{A3}.

Thus, for cascades~(1),\,(5), produced by the initially 
pure $B_s$ or $\overline B_s$ states we obtain
\begin{equation}
W^+_s(t_1,t_2) = \left|a_s \frac{1+\epsilon_K}{1-\epsilon_K}
\right|^2
\cdot F(t_1,t_2; \lambda_s, -1),\;\;\;\;
W^-_s(t_1,t_2) = |a_s|^2 \cdot F(t_1,t_2; \lambda_s, 1);\,\,\,\,
\end{equation}
\begin{equation}
\,\,\,\overline W^-_s(t_1,t_2) = \left |\overline a_s 
\frac{1-\epsilon_K}{1+\epsilon_K}\right |^2 \cdot 
F(t_1,t_2; \lambda_s^{-1}, -1),\,\,\,\,
\overline W^+_s(t_1,t_2) = |\overline a_s |^2\cdot F(t_1,t_2; 
\lambda_s^{-1}, 1).
\end{equation}
The time distributions for the cascades~(1),\,(6) are
$$W_s^{\pi \pi}(t_1,t_2) = \frac{|a_s|^2}{2}\cdot
\frac{1+|\epsilon_K|^2}{|1-\epsilon_K|^2} \cdot
F(t_1,t_2; \lambda_s, -\eta)\,,$$
\begin{equation}
\overline W_s^{\pi \pi}(t_1,t_2) = \frac{|\overline 
a_s|^2}{2}\cdot
\frac{1+|\epsilon_K|^2}{|1+\epsilon_K|^2} \cdot
F(t_1,t_2; \lambda_s^{-1}, \eta)\,.
\end{equation}
Here $\eta=\eta_{\pi\pi}$~ is the standard parameter 
describing the contribution of $K_L$ to the decay~(6).
The function $F$ has a rather complicated structure. If 
we are mainly interested in the primary-beauty decay 
distribution, it may be presented as
$$F(t_1,t_2;\lambda_s, c) = \exp (-\Gamma_+^{(s)}t_1)
\cdot A(t_2;\lambda_s,c) + \exp(-\Gamma_-^{(s)}t_1)\cdot 
A(t_2; -\lambda_s,c) \;\;\;\;\;\;\;\;\;\;\;\;\;\;\;\;\;$$
\begin{equation}
\;\;\;\;\;\;\;\;\;\;\;\;\;\;\;\;\;\;\;\;+ 2 \exp(-
\Gamma^{(s)}t_1)
\cdot {\rm Re}\,[\exp(- i\Delta m_s t_1) \cdot 
B(t_2;\lambda_s, c)]
\end{equation}
with $\Gamma^{(s)} = (\Gamma_+^{(s)} + \Gamma_-^{(s)})/2\,,\,\,
\Delta m_s = m_-^{(s)} - m_+^{(s)}$;\,\, $m_{\pm}^{(s)}$ 
and $\Gamma_{\pm}^{(s)}$ are the masses and widths of two 
beauty-strange meson eigenstates. We define $B_+^{(s)}$ as 
the approximately $CP-$even state which is the main source 
of $J/\psi K_L$, while the approximately $CP-$odd state 
$B_-^{(s)}$ is the main source of $J/\psi K_S$ (see 
refs.~[11, 13--15] for more detailed discussions related 
to this point; see also the concluding discussion below). 
$\Delta m_s$ is defined here so as to have a positive kaon 
analog $\Delta m_K = m_L - m_S$. The coefficients $A$ and 
$B$ themselves have a similar three-term  structure: 
$$ A(t_2;\lambda_s, c) = \exp(-\Gamma_St_2)\cdot \left|
\frac{1+\lambda_s}{4}\right|^2 + \exp(-\Gamma_Lt_2)\cdot \left|
c \cdot\frac{1-\lambda_s}{4}\right|^2\,\,\,\,\,\,\,\,\,\,\,\,\,
\,\,\,\,\,\,\,\,\,\,\,\,\,\,\,\,\,\,\,\,\,\,\,\,\,\,\,\;\;\;\;$$
\begin{equation}
\;\;\;\;\;\;\;\;\;\;\;+ 2 \exp\left(-\frac{\Gamma_S + \Gamma_L}
{2}t_2\right) \cdot {\rm Re}\,\left[\exp(-i\Delta m_K t_2)
\cdot c \cdot \frac{1+\lambda^\ast_s}{4}\cdot\frac{1-\lambda_s}        
{4}\right] ; 
\end{equation}
$$ B(t_2;\lambda_s, c) = \exp(-\Gamma_St_2)\cdot \frac
{1+\lambda^\ast_s}{4}\cdot\frac{1-\lambda_s}{4} +
 |c|^2\cdot\exp(-\Gamma_Lt_2)\cdot\frac{1-\lambda^\ast_s}{4}
\cdot \frac{1+\lambda_s}{4}\;\;\;\;\;\;\;\;\;\;\;\;\;\;\;\;$$
\begin{equation}
+\exp \left(-\frac{\Gamma_S+\Gamma_L}{2}t_2\right)\cdot
\left[c\cdot\exp(-i\Delta m_Kt_2)\left|\frac{1+\lambda_s}{4}
\right|^2 +c^\ast\cdot\exp(i\Delta m_Kt_2)\left|\frac{1-              
\lambda_s}{4}\right|^2\right]\,,          
\end{equation}
where $\Gamma_{S,L}$ are the widths of $K_{S,L}$.

Expressions (12)--(14) show that time-distributions of the 
primary and secondary decays are essentially correlated. 
Information about the properties of decays and mixing of $B_s$ 
and $\overline B_s$ continue to be encoded in the time-dependence 
of the daughter kaon decays (on $t_2$), even when all $B_s-$decay 
times (i.e., $t_1$) have been integrated over (e.g., because of 
insufficient resolution). The resulting time distributions of 
secondary decays can be easily expressed through an integral 
consisting again of three terms with different $t_2$-dependences: 
$$I(t_2;\lambda_s,c)=\int_{0}^{\infty}dt_1\,F(t_1,t_2; 
\lambda_s,c) = \exp(-\Gamma_St_2)\cdot C(\lambda_s) +  |c|^2 
\cdot\exp(-\Gamma_Lt_2)\cdot C(-\lambda_s) 
\,\,\,\,\,\,\,\,\,\,\,           
$$
\begin{equation}
\,\,\,\,\,\,\,\,\,\,+2\exp \left(-\frac{\Gamma_S+\Gamma_L}
{2}t_2\right)\cdot {\rm Re}\,[c\cdot\exp(-i\Delta m_Kt_2)
\cdot D(\lambda_s)]\,;
\end{equation}
\begin{equation}
4\Gamma^{(s)}C(\lambda_s)=[(1+|\lambda_s|^2)/2-y_s{\rm Re}
\lambda_s] (1-y_s^2)^{-1}+[(1-|\lambda_s|^2)/2-x_s
{\rm Im}\lambda_s] (1+x_s^2)^{-1},                                    
\end{equation}
\begin{equation}
4\Gamma^{(s)}D(\lambda_s)=[(1-|\lambda_s|^2)/2+iy_s{\rm Im}
\lambda_s] (1-y_s^2)^{-1}+[(1+|\lambda_s|^2)/2-ix_s
{\rm Re}\lambda_s] (1+x_s^2)^{-1}.
\end{equation}
\begin{equation}
y_s = (\Gamma_+^{(s)}-\Gamma_-^{(s)})/2\Gamma^{(s)}\,,\;\;
\;\;\;\;x_s =\Delta m_s/\Gamma^{(s)}\,.
\end{equation}
If the main goal is to extract $\Delta m_s$, then in the 
frame of the Standard Model one may neglect $CP-$violation 
and to a good approximation use $\lambda_s=-1$, thus strongly 
simplifying the above expressions. The expectations 
$$|y_s|\ll 1\;\; {\rm and} \;\; |x_s|\gg 1$$
give further simplifications; e.g., time distributions of
secondary leptons with charge $\pm1$ [or of cascading $K^0(
\overline K^0)\rightarrow \pi^+\pi^- $ decays] for initially 
pure $B_s$ are determined by $I(t_2;-1,\mp1)$ [or by 
$I(t_2;-1,-\eta)$] and take the form
\begin{equation}
W_s^{\pm}(t_2) \propto (1+y_s)\exp(-\Gamma_St_2)+
(1-y_s)\exp(-\Gamma_Lt_2) \mp\frac{2}{x_s}\sin(\Delta m_Kt_2)
\exp\left(-\frac{\Gamma_S +\Gamma_L}{2}t_2\right);
\end{equation}
$$W_s^{\pi\pi}(t_2) \propto (1+y_s)\exp(-\Gamma_St_2)+
(1-y_s)|\eta|^2\exp(-\Gamma_Lt_2)\;\;\;\;\;\;\; $$
\begin{equation}
\;\;\;\;\;\;\;-\frac{2}{x_s}|\eta|\sin(\Delta m_Kt_2 - \varphi)
\exp\left(-\frac{\Gamma_S +\Gamma_L}{2}t_2\right)
\end{equation}
with $\varphi = \arg \eta$.

These distributions look like distorted time distributions 
for the corresponding modes of the usual  kaon decays. 
The distortion of non-oscillating terms reveals the width 
difference (for $B_s$ !), while the amplitude of oscillations 
directly depends on the $B_s$ mass difference. We see that the 
amplitude decreases with increasing $\Delta m_s$, but 
slower than other integrated effects. The sign of the 
oscillating term is directly related to the sign of the 
mass difference. In this respect eqs. (19), (20) are 
similar to the corresponding results for cascading 
decays of $B_d$~\cite{ARS}.

In conclusion we briefly discuss the identification method
for $B-$meson eigenstates applied above (as well as in 
refs.~[9--11, 16]) and compare it to other methods used
in the literature. The starting point is that detailed 
studies of an unstable particle (especially, short-living)  
are possible only through observation of its decay final states. 
So if $CP-$conservation were exact, our definition of 
$CP-$parity for the eigenstates and the method of their 
identification by $CP-$parity of their decay final states 
would be exact and quite adequate. In the case of really 
small $CP-$violation our definition of the $CP-$parities 
becomes approximate, but still unambiguous and independent 
of a decay mode. When the intrinsic $CP-$violation increases, 
some decay modes may give indefinite (or even reversed) 
$CP-$parities to the eigenstates. So, our definition of 
$CP-$parities of the eigenstates may appear mode-dependent.
Nevertheless, for most decay final states our separation 
of the eigenstates continues to be unambiguous for each 
particular decay mode (even if the eigenstates' $CP-$parities 
become reversed). Necessary comparisons of various decay modes
are possible by comparing the signs of $\Delta m$ and $\Delta
\Gamma$ as measured in the corresponding decay modes. This 
problem is very interesting by itself. It is even more so 
if $CP-$violation is generated by the CKM-mechanism, since
the known information on the CKM-matrix makes very probable 
such large intrinsic violation (we mean that, according to 
experimental data, at least two angles of the unita\-rity 
triangle for the $b-$quark tend to be large, and some final 
states could produce inverted $CP-$parity in comparison to 
others, if $\pi/4 < |\gamma| < 3\pi/4$; see discussion 
in ref.~\cite{Dun}).       

Identification of nearly degenerate states by their heavier
or lighter  masses, denoted usually by $H$ or $L$ indices,
is universal (i.e., definitely mode-independent) and seems at 
first sight more natural and convenient. But experimentally it 
is inapplicable directly, even to neutral kaons. Classification 
of kaon eigenstates by their masses became meaningful only as a 
result of complicated interference experiments where the sign of 
$\Delta m_K$ was measured. 

Identification of eigenstates by their longer ($L$) or shorter 
($S$) lifetimes is good for kaons since their lifetimes differ 
very strongly (more than 500 times) and at least $K_L$ (but not
$K_S$) may be easily separated. However, such a procedure is 
definitely useless experimentally for $D$ and $B_d$ mesons 
(and hard enough for $B_s$). 

Contrary to these two, the above procedure is well-defined for 
all particular experimental conditions and looks natural in 
theoretical calculations for any particular decay mode. After 
measuring the signs of $\Delta m$ and $\Delta\Gamma$ all the 
procedures should become equivalent.

The technique outlined here for measuring very large $B_s - 
\overline B_s$ mixing effects does not require fine resolution 
in the $B_s-$decay time. Instead it requires copious production 
of $B_s-$mesons that should be flavor-tagged. The mode $B_s 
\rightarrow J/\psi\overline K^0$, although CKM suppressed, is 
triggerable at hadron accelerators. If $x_s$ remains out of 
reach for existing  vertex technology, we recommend detailed 
feasibility studies of the cascading processes $B_s(t)\rightarrow 
J/\psi K^0 [\rightarrow\pi^+\pi^-, \pi l\nu]$. Depending on 
the detector configuration, one may 
wish to cut on $B_s$ decay times that are very close to the 
primary interaction vertex. Although the detector may not be 
able to resolve the rapid $(\Delta m_st)$-oscillations, it 
may be able to give very interesting information, e.g., to
distinguish the two different $B_s$ lifetimes. Because 
integrating over all $t_1$ removes such information, we 
presented in detail (see Eqs.(9)--(14)) the complete time 
distributions of the cascade processes with which further 
detailed analyses can be conducted.

This note addressed only one aspect of coherence effects between 
heavy neutral mesons and their decay final states containing a 
single neutral kaon. Detailed investigations of such coherence 
effects may prove useful also in measuring a small lifetime 
difference for the two $B_d$ eigenstates, and in measuring 
small $D^0 -\overline D^0$ mixing parameters \footnote{The 
complication due to doubly Cabibbo suppressed modes can be 
incorporated in a straightforward fashion. It will be presented 
elsewhere.}. Furthemore, they will shed important light upon 
$CP-$violation and extraction of CKM parameters. We hope to 
return to those intriguing issues in the future. 
 
\section*{Acknowledgements}

We thank D.Richards for reading the manuscript. Ya.Azimov 
appreciates the hospitality of the Theory Group at Fermilab 
where this work was done. This work was supported in part 
by Universities Research Association Inc., under Contract
No. DE-AC02-76CH03000 with the United States Department of
Energy.

\end{document}